\documentclass[twocolumn,aps]{revtex4}
\usepackage{amsmath}
\usepackage{epsfig}
\newcommand{\f}{\begin{equation}}
\newcommand{\ff}{\end{equation}}

\newcommand{\beq}{\begin{equation}}
\newcommand{\eeq}{\end{equation}}
\newcommand{\be}{\begin{equation}}
\newcommand{\ee}{\end{equation}}
\newcommand{\bea}{\begin{eqnarray}}
\newcommand{\eea}{\end{eqnarray}}

\begin{document}

\title{Holography and the  scale-invariance of
density fluctuations}

\author{Jo\~{a}o Magueijo$^{1,2,3}$, Lee Smolin$^{1,4}$ and Carlo~R.~Contaldi$^3$}

\affiliation{$^1$ Perimeter Institute for Theoretical Physics, 31
Caroline St N, Waterloo N2L 2Y5, Canada\\
$^2$ Canadian Institute for Theoretical Astrophysics, 60 St George
St, Toronto M5S 3H8, Canada\\
$^3$ Theoretical Physics Group, Imperial College, Prince Consort
Road, London SW7 2BZ, UK\\
$^4$ Department of Physics, University of Waterloo,
Waterloo, Ontario N2L 3G1, Canada}

\begin{abstract}
We study a scenario for the very early universe in which there is
a fast phase transition from a non-geometric, high temperature
phase to a low temperature, geometric phase described by a
classical solution to the Einstein equations. In spite of the
absence of a classical metric, the thermodynamics of the high
temperature phase may be described by making use of the
holographic principle.  The thermal spectrum of fluctuations in
the high temperature phase manifest themselves after the phase
transition as a scale invariant spectrum of fluctuations.  A
simple model of the phase transition confirms that the near scale
invariance of the fluctuations is natural; but the model also
withstands detailed comparison with the data.

\end{abstract}

\pacs{PACS Numbers: *** }


\date{\today}

\maketitle In this letter we propose a new hypothesis about the very
early universe in which the notion of holography plays a key role.  We
will see that it gives rise to a scale invariant distribution of
density fluctuations, without invoking inflation.  This hypothesis is
inspired by a scenario that has been proposed
recently~\cite{non-geometry} in which the universe begins in a high
temperature phase (Phase I) which is not described in terms of fields
on classical spacetime manifolds; rather it has a non-geometrical,
purely quantum mechanical description, which can be expressed simply
in terms of the holographic principle. The spacetime geometry is
created in a phase transition, into Phase II, in which it is
appropriate to describe the universe to a decent approximation in
terms of fields or other degrees of freedom (particles, strings, etc)
moving in a classical background geometry. (This may be called {\it
geometrogenesis}). The phase transition happens at temperature $T_c$
and is abrupt (in a sense to be made precise later) and imprints on
Phase II a spectrum of thermal fluctuations which arise in Phase I.

The innovation in this paper is to use the holographic principle to
characterize the two phases.  That principle, as enunciated by 't
Hooft~\cite{thooft}, asserts that {\it any} region of space bounded by
a surface $\cal S$ of area $A$ can be described by a finite number of
degrees of freedom given by $N= {A}/(4G\hbar)$.
These evolve according to a fundamental dynamics, given by Hamiltonian
$H$.  The term ``holography'' has also acquired other
different meanings~\cite{susk}, but we stress that we use it in its
original sense~\cite{otherholo}.

To describe classical physics there must be non-local correlations
among the degrees of freedom on the surface, so as to make it appear
that the dynamics is local in a volume $V$ described by a classical
geometry in a region $\cal R$ that $\cal S$ bounds.  Close to the
ground state we expect that $V \approx A^{\frac{3}{2}}$ whenever the
curvature can be neglected.
But in Phase I there is not yet a classical spacetime geometry. We
propose that this phase be then characterized as an disordered phase,
which means that the non-local correlations which are needed on the
surface to construct the illusion of local three dimensional physics
are not yet established, and are peculiar to Phase II.  At the same
time, the degrees of freedom in Phase I are highly interconnected, as
in the model of \cite{non-geometry} and so reach equilibrium before
the phase transition.  This solves the horizon problem, as noted in
\cite{non-geometry}; the next challenge is to generate an almost scale
invariant spectrum of fluctuations.

Our idea is very simple, and even though we'll derive it step by step
later, we sketch it first.
If Phase I is in thermal equilibrium we expect that the $N$ (local)
fundamental degrees of freedom are excited to energy $T$.  Thus, in
this disordered phase the holographic principle implies that
$E= NT =  \frac{A}{4\hbar G}T$;
that is, for $T > T_c$ the specific heat at fixed ${\cal R}$
is  $c_{\cal R}={\partial E\over \partial T}= N$,
which scales like the area.
A well known result tells us that the thermal fluctuations in the
energy contained in a fixed region are given by $\sigma^2_E
= T^2 c_ {\cal R}$. Therefore in Phase I we have $\sigma^2_E
\approx {T^2\over 4\hbar G} A$, and because this scales like $A$
(instead of $V$, as usual) we have a scale-invariant spectrum of
fluctuations, rather than the usual white noise.
If the phase transition is ``fast'', when the classical metric emerges
in Phase II these fluctuations propagate to the potential. Most modes
are now outside the Hubble radius, so we end up with a scale-invariant
spectrum of adiabatic density fluctuations, with amplitude fixed by
the ratio $T_c/T_{Pl}$.

We shall proceed as follows.  We first present a description of the
two phases, justified by basic facts of the quantum theory of gravity,
and compute the associated density fluctuations.  We next propose a
scenario for the generation of classical geometry during the
transition. This results in a scale invariant distribution of fluctuations outside
the horizon of the classical geometry that emerges at the end of the
transition.  We then devote some time to weakening our hypotheses,
showing how the essential results apply to a large class of models of
this type. Finally we quantify departures from scale-invariance,
predicting the amplitude, spectral index, and its running in terms of
$T_c$, and $\gamma$, the critical exponent characterizing the phase
transition. A comparison with other models and a word on tensor modes
closes this Letter.

To characterize the two phases in more detail we must make a few general assumptions.
Firstly, that
the universe can be described at all times as having fixed three dimensional
spatial topology, but not necessarily a fixed classical metric.
Secondly, that the physics in {\it any}  region  $\cal R$  can be described in a hamiltonian
formulation  in terms of a Hilbert space ${\cal H}_{\cal R}$.
$\cal R$ has a boundary $\partial { \cal R} = {\cal S}$.
And finally that
among the operators in ${\cal H}_{\cal R}$ are $\hat{A}$,  the area of
${\cal S}$, the hamiltonian constraint $\cal C$ and diffeomorphism
constraints $\cal D$ and a boundary contribution to the hamiltonian
$\hat{h}= \int_{ \cal S} \hat{\mu} $ where $\hat{\mu}$ is a local
operator on the boundary.  The total hamiltonian for quantum spacetime
and matter in $\cal R $ is then,
\f
\hat{H} =  \int_{\cal S} \hat{\mu} + \int_{\cal R} N {\cal C} +
v^a {\cal D}_a
\ff
The quantum states of interest are physical states that are in the kernel of $\cal C$ and $\cal D$.  These assumptions are common
to many approaches to quantum gravity as they follow just
from diffeomorphism invariance.

We then hypothesize that there are two different kinds of solutions to
the quantum constraints which each characterize one of the phases.  We
characterize Phase II as being {\it ordered three dimensionally.} This
means that  there is a classical non-degenerate  three metric $q_{ab}$
on  $\cal  R$  such that  the  physics  can  be  described to  a  good
approximation by a semi-classical state built from $q_{ab}$.
We characterize Phase I as being {\it disordered}.  This means that
there is no such three dimensional classical metric $q_{ab}$.

In phase II there is no mass gap, so that, in the limit of infinite
area, there is a continuum of states above a ground
state $ | {\rm II} \rangle $
where $\hat{H} |{\rm II}\rangle =0$.  These correspond to gravitons and other
massless excitations.  In this phase correlations develop on the
boundary $\cal S$ corresponding to the fact that the lowest energy
excitations have long wavelength.

In phase I there is a mass gap, of order of the Planck mass,
$M_{Pl}$. (This is a property of one form of the $ADM$ energy in
LQG given by~\cite{Thiemann}.) Thus, this is a high temperature
phase, with $T \approx M_{Pl}$. We hypothesize that in this phase
there are no correlations on the boundary. At the same time
quantum geometry is quantized on the boundary.  Thus, let $\sigma$
and $\sigma^\prime $ be regions of the boundary $\cal S$ and let
$\hat{h}( \sigma) = \int_\sigma \hat{\phi}$ where $\hat\phi$ is
the degree of freedom on the boundary. Thus, ${\rm Tr} [
\rho_T^{\rm I}   \hat{h}( \sigma) \hat{h}( \sigma^\prime ) ] =0$
for $\sigma \cap \sigma^\prime = 0$, suggesting that \be {\langle
E\rangle} = {\rm Tr} [  \rho_T^{\rm I}  \hat{H} ]  = b M_{Pl}^2 T
{\langle A\rangle} \ee where $b$ is some dimensionless constant
and ${\langle A\rangle} = {\rm Tr} [  \rho_T^{\rm I}  \hat{A} ] $,
so that {\it in Phase I energy is proportional to the energy on
the boundary. } This implies in particular that the specific heat
at fixed area is proportional to the area: \f c_{A}=
{\left(\frac{\partial {\langle E\rangle}}{\partial
T}\right)}_{{\langle A\rangle}} = b \frac{{\langle
A\rangle}}{\hbar G} \label{key1} \ff Since we have a Hamiltonian
on the surfaces in phase I we can study its thermodynamics. Here
we are allowed to use only the following quantities which are
assumed to exist for the region $\cal R$:  i) a Hilbert space of
spatially diffeomorphism invariant states;  ii) an area operator
$\hat{A}$;  iii) a hamiltonian operator $\hat{H}$ as described
above. The thermal physics is defined by the  partition function:
\begin{equation}
Z={\sum_r} e^{-\beta E_r} \ ,
\end{equation}
where $\beta = T^{-1}$. The total energy $U$ inside region ${\cal R}$ is:
\begin{equation}
U={\langle E\rangle}={{\sum _r} E_r e^{-\beta E_r} \over {\sum_r}
e^{-\beta E_r}}=-{d\log Z\over d\beta}
\end{equation}
and its variance by
\begin{equation}\label{varE}
\sigma^2_E={\langle E^2\rangle}-{\langle E\rangle}^2={d^2\log
Z\over d\beta^2}= -{dU \over d\beta}=T^2c_{{\langle A\rangle}}
\end{equation}
where $c_{{\langle A\rangle}}$ is the specific heat at constant expectation value of area, ${\langle A\rangle}$. We use (\ref{key1}) above to write
\f
\sigma^2_E = b T^2 {\langle A\rangle}
\label{key2}
\ff
and this is all we need from Phase I.

In phase II we have, in addition to the observables of Phase I,
the metric in the interior of the region $\cal R$, as well as all
other usual quantities. Regarding the transition from Phase I to
II we make the following hypotheses:

A) The phase transition begins when the temperature falls to a
critical temperature $T_c$.

B) The phase transition proceeds from large scales down to a
small scale $l_0$. At any $T<T_c$ the geometry is characterized by
a length  scale $R(T)$ such that the geometry appears classical to
all  modes of the field  with wavelength $\lambda  \geq R(T)$. The
geometry that those  modes probe should be as  simple as possible,
hence, up  to small fluctuations imprinted  on  it by the  density
fluctuations created in  Phase I, it is homogeneous and isotropic.
$l_0$ is not zero because if we probe small enough even in the
ground state the continuum dissolves into the quantum geometry.

C) Since $R(T)$ is infinite at $T=T_c$ and then falls to $l_0$, 
the dependence of $R$ on $T$ during the transition is
modelled by a function \be \label{gamma} {R(T) \over
l_0}={\left(T_c\over T_c-T\right)}^\gamma \ee valid for $T\le
T_c$. This introduces a parameter which is the critical exponent
$\gamma$.

Our purpose is now to compute the spectrum of fluctuations left
over in Phase II. We recall that in Phase I there is a Hamiltonian
but not a stress-energy tensor: therefore we can only talk about
energy fluctuations. Also there is no sense of Fourier modes, so
only the variance $\sigma^2_E$ can be defined. We must now work
out onto which Phase II structures we should map these energy
fluctuations, as they emerge out of phase I. Firstly it's clear
that as the concept of length and volume are created we'll have
${\langle A\rangle}=A=4\pi R^2$, and $V(R)={4\over 3}\pi R^3$.
This defines energy density perturbations $\delta \rho = \delta
E/{V}$. We then have $\sigma^2_\rho(R)={1\over R^6}\sigma^2_E(R)$,
where $\sigma^2_\rho(R)$ is the mean square perturbation in the
region. We also have $c_A=c_V$ (fixed ${\cal R}$ now means fixed
${\langle A\rangle}$ and ${\langle V\rangle}$), so using
(\ref{key1}) and ({\ref{varE}) we have \f \label{sigma2rho}
\sigma^2_\rho(R)={\sigma^2_E(R)\over R^6}={T^2\over R^6}c_V =
\frac{4 \pi b T^2 }{G R^4} \ff These fluctuations emerge in phase
II outside the horizon defined by $H^{-1}\sim T_c^{-2}$; 
we assume that they are mapped into a $\delta\rho$ defined in the
longitudinal/comoving gauge (see~\cite{liddle,ks}). 
Also other 
components of the stress energy tensor are now well defined: we set
the anisotropic stress to zero, and assume a pressure fluctuation
so as to make the fluctuations adiabatic (this is convenient but
not strictly necessary). Then the metric in Phase II may be written
$ds^2 = a^2 (\eta ) [-d\eta^2( 1-2 \Phi ) + (1+2 \Phi ) dx^2]$,
where $\Phi$ labels small fluctuations in the metric, and 
these are related to the comoving $\delta\rho$ by the
Poisson equation 
\be
k^2\Phi=4\pi G a^2\delta\rho
\ee
{\it for all modes}, large and small (again see~\cite{liddle,ks}). 
If we relate the
density fluctuations defined by (\ref{sigma2rho}) to their
dimensionless power spectrum by the formula, $\sigma_\rho^2(R)
\sim {\cal P}_{\delta \rho} (k={a}/{R})$ (see~\cite{liddle}), we
finally get
 \be {\cal
P}_\Phi (k) =\frac{16\pi^2G^2a^4}{k^4} {\cal P}_{\delta\rho}  (k)
\sim \frac{ G^2 a^4}{k^4}{\sigma^2_\rho{\left(R={a/ k}\right)}}.
\ee If we neglect the variation in $T$ during the transition,  we
find using (\ref{sigma2rho}), \f \label{calP0} {\cal
P}_\Phi={k^2\over a^2}T^2 {c_V}\sim
 \frac{G}{\hbar} T_c^2  = \frac{T_c^2}{T_{Pl}^2}
\ff that is a  scale-invariant spectrum, with amplitude ${\cal
A}\sim T_c/
 T_{Pl}\sim 10^{-5}$. Now, in fact, given (\ref{gamma}), the spectrum in phase II cannot
be exactly scale-invariant, because different scales freeze-in at
slightly different temperatures and therefore with slightly
different amplitudes. Using  (\ref{gamma}) we find that more
precisely \be \label{powern01} {\cal P}_\Phi(k)=2 \left(
\frac{T_c}{T_{Pl}}\right)^{2} {\left [1-{\left(l_0 k\over
a_c\right)}^{1\over \gamma} \right]}^{2} \ee
(we have  neglected the variation in $a$ during the phase transition so that $k\approx
 a_c/R(T)$).
Since the spectrum is slightly red we believe we are not affected
by the concerns voiced in~\cite{joy,dav}.

This interesting result does not depend on all the assumptions on
Phase I made above. Absence of metric or vanishing of $V$ in Phase I,
for example, are not strictly needed. Indeed {\it the only requirement
for scale-invariance is that the specific heat for a fixed region be
proportional to the area}.
This requirement is realized in any other model where energy behaves like a ``surface tension'',
that is, $E\propto R^2 e(T)$, with general $e=CT^\zeta$.
For our model $\zeta=1$ and $C\sim 1$, but if there were no scale
(like $G$) in the problem
we would expect $e\propto T^3$, just like $E=VT^4$ usually.
In these alternative scenarios our results only change in detail. For example
we would get
\be
\label{powern0}
{\cal P}_\Phi(k)=C \zeta \left( \frac{T_c}{T_{Pl}}\right)^{\zeta+1} {\left [1-{\left(l_0 k\over
a_c\right)}^{1\over \gamma} \right]}^{\zeta+1}
\ee
so $T_c\approx {10^{-{10\over \zeta+1}}}{T_{Pl}\over C\zeta}$.
If  $C$ is not order 1, and if $\zeta$ is not 1, $T_c$ may be quite
different from our estimate, but everything else only changes in detail.
In what follows we shall explore $\zeta=1,3$ but set $C=1$.

Our results  are also valid if  the energy remains  extensive but {\it
the thermal correlations are  string-like infinite tubes} (rather than
volumes with diameter $\sim  1/T$), with a section $\Sigma=\Sigma(T)$.
Thermal fluctuations may  be seen as a Poisson  process (with variance
$\sigma^2\sim N$) for these uncorrelated regions. Usually their number
$N$ scales  like the volume  ($N=VT^3$); thus a white  noise spectrum.
But  for filamentary  correlated  regions, $N$  scales  like the  area
($N\sim V/(R\Sigma)\sim A/\Sigma(T)$), i.e.  scale-invariance. This is
another way  of understanding our  derivations. It is also  the reason
why   the    Hagedorn   phase   scenario    of~\cite{rob}   leads   to
scale-invariance, and it may connect with the work of~\cite{renata}.

We now relate the breaking of scale invariance to observation.
Referring the non-scale-invariant bit of the spectrum to pivot $k_\star=0.002$~Mpc$^{-1}$ as usual~\cite{WMAP03cosmo} we thus get
\be
\label{powern}
{\cal P}_\Phi (k)=\zeta \left( \frac{T_c}{T_{Pl}}\right)^{\zeta+1} {\left
[1-{\left( \alpha{k\over k_\star}\right)}^{1\over \gamma}
\right]}^{\zeta+1}
\ee
with
\be
\alpha={l_0\over a_c}k_\star
\approx 5.5 \times 10^{-29}{l_0\over l_P}{T_c\over T_{Pl}} .
\ee
(we have ignored variations in the $g$ factor relating
$T$ and $1/a$.)
The term in $\alpha$ is therefore small and so (\ref{powern}) can
be expanded into a scale invariant term plus
and a negative, blue component. The total is therefore a red spectrum with effective tilt
\be
\label{ns}
 n-1={d\ln {\cal P}\over d\ln k}=-{\zeta +1\over
\gamma} {\left( \alpha{k\over k_\star}\right)}^{1\over \gamma}
\ee
Note that we get a red tilt to the spectrum because we postulated
that  the phase transition is outside-in.

Observations probe at most $\ln(k/k_\star)\sim 10$, so
(\ref{ns}) can also be expanded, leading to
a series $n=n_0+n_1+n_2+...$, with  $n_0=1$,
\be
n_{1}=-{\zeta+1\over \gamma}\alpha^{1/\gamma}
\ee
and most crucially the second order prediction
\be
\label{consist}
 {dn_2\over
d\ln k}={n_{1}\over \gamma} \quad {\rm i.e.}\quad
 {dn\over
d\ln k}\approx  {n-1\over \gamma}
\ee
which can be seen as a ``consistency condition''.

Given the smallness of $\alpha$, departures from scale invariance are
usually very small. If $l_0\sim T_c^{-1}$ they are maximized (at scale
$k_\star$) for $\gamma=-\ln\alpha\approx 65$, with $n-1\approx 0.005
(\zeta +1)$, of the order of a few percent. However, if $\gamma\sim 1$
the deviations would be of order $10^{-30}$. For $l_0\gg T_c^{-1}$ we
can generate larger deviations, but still only for $\gamma\sim
-\ln\alpha\approx 65 -\ln (l_0 T_c)$; with $\gamma\sim 1$ again
leading to infinitesimal deviations.

Using a modified version of the {\textsc CAMB} code \citep{Lewis:2000}
we calculate the predicted CMB and Large Scale Structure (LSS) power
spectra for our model. We then use a Monte Carlo Markov Chain (MCMC)
for sampling the likelihood as a function of model
parameters~\citep{Lewis:2002ah}. We fit a conventional combination of
CMB and LSS data sets~\cite{data}.  We adopt the same number (six) of
parameters as conventional (flat, power law) models, but trade $n$ and
amplitude $A$ parameters for $\gamma$ and $T_c$.  The fits obtained
are therefore directly comparable to the standard power law
$P_\Phi(k)$ based fits since the number of model parameters are the
same. The results are shown in Fig.~\ref{fig1} for $\zeta=1,3$ and
$l_0^{-1}=T_c,10^{11}$ GeV. We find that all runs give good fits to
the data although the choice of $\zeta=1$ requires very large values
of $\gamma$ to achieve a sufficient amount of red tilt.  However for
$\zeta=3$ we obtain a slightly better fit that the conventional power
law models, as detailed in Table~\ref{tab:results}.

\begin{figure}
\centering
\includegraphics[width=8cm,height=8cm]{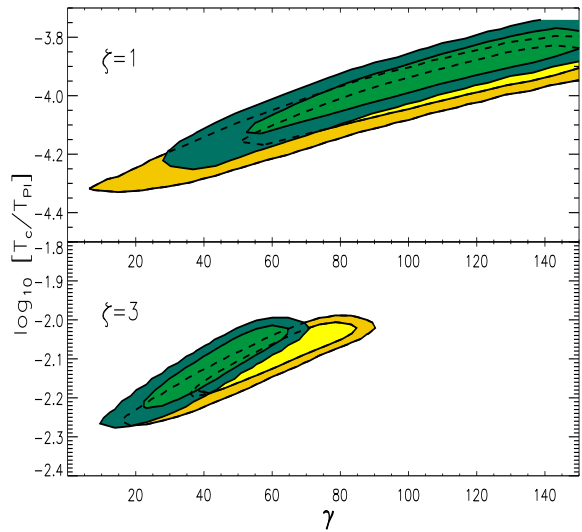}
\caption{Marginalized confidence contours for the $\gamma$ and
$\ln[T_c/T_{Pl}]$.  The contours represent the 68\%
and 95\% integrals. The top
and bottom panels show results for $\zeta=1$ and $3$,
respectively. The underlying (yellow) region is for $l_0=T_c^{-1}$
while the overlayed (green) region is for $l_0^{-1}=10^{11}$GeV. }
 \label{fig1}
 \end{figure}

\begin{table}
\begin{tabular}{cc||ccc}
\hline\hline \rule[-2mm]{0mm}{6mm} $l_0$&$\zeta$&
$\frac{T_C}{T_{Pl}}(\times 10^4)$ & $\gamma$ & $\Delta \ln\,
L$\\ \hline\hline $ T_c^{-1}$& $\zeta=1$ & $1.04^{+0.13}_{-0.14}$ & $
103.7^{+ 46.2}_{- 42.4} $ &$-0.75$\\ $ T_c^{-1}$& $\zeta=3$ &
$80.3^{+6.0}_{-6.4}$ & $ 59.9^{+ 8.0}_{- 8.8} $ &$+0.04$\\ $
(10^{11}\mbox{ GeV})^{-1}$& $\zeta=1$ & $1.18^{+0.16}_{-0.16}$ & $
104.2^{+ 45.6}_{- 39.5} $ &$-0.32$\\ $ (10^{11}\mbox{ GeV})^{-1}$&
$\zeta=3$ & $76.4^{+6.6}_{-6.2}$ & $ 43.0^{+ 6.6}_{- 6.8}$
&$+0.07$\\ \hline
\end{tabular}
\caption{Median values  for $\gamma$ and $T_c$.
The errors are obtained from the 68\% confidence
integrals. The $\Delta \ln\, L$ is with respect to the best
conventional power law spectrum.}
\label{tab:results}
\end{table}

It is interesting to compare our model with other alternatives.
Slow-roll inflation also predicts a near scale invariant spectrum,
i.e. $n_0=1$, with correction $n_1=-6\epsilon+2\eta$, in terms of
slow-roll parameters $\epsilon={1\over 2}{\left( V'/ V\right)}^2$ and
$\eta = {V''/ V}$, (where $V(\Phi)$ is the inflaton potential).
There is also a second order logarithmic running of $n$, but for a general
$V$ this is independent of $n_1$, since it depends
on $\xi^2={V' V'''/ V^2}$. More generally inflation can produce {\it any}
spectrum of scalar fluctuations if one carefully designs $V(\phi)$,
and it's not a falsifiable theory until we consider the tensor modes,
which do impose a consistency condition. By contrast we predict a scalar
``consistency condition'' (\ref{consist}).

Our work has obvious parallels with that of~\cite{rob} on the Hagedorn
phase.  However, our phase transition is a transition in the
description of space-time rather than in the matter content. This
profound conceptual difference has a very practical implication: the
function (\ref{gamma}) controlling the progress of the phase
transition is entirely different from that (Eqn.~(\ref{calP0}))
controlling the amplitude of the fluctuations. Thus the naturalness of
scale-invariance. This is not the case with~\cite{rob} where both
scales are controlled by $T_c/(T_c-T)$, and a degree of (debatable)
fine tuning is required.  It would be interesting to study how those
models fare with deviations from exact scale-invariance.
Our scenario can also be seen as a variant of a varying speed
of light cosmology~\cite{vsl0} in that in Phase I all degrees of
freedom are assumed to be in causal contact and thermal equilibrium.

To conclude and summarize, we have presented a model for the
emergence of classical space-time from a quantum, non-geometric
pre-era informed by  a version  of the  holographic principle. Our
model  develops the general  idea  in  \cite{non-geometry}  in
that  we  make  particular hypotheses  about the  phase
transition,  which we  labelled A)  to C) above.  These
hypotheses could be  checked in the context of different models of
quantum gravity.  What  is remarkable  is  that there  are
specific consequences of  non-perturbative quantum gravity models
that may be calculable which, given our picture, map on to
quantities which are  measurable  in  the  CMB.  In  particular
the  phase transition temperature   is  mapped   to  the
amplitudes  of fluctuations  by (\ref{calP0}), the  direction of
the transition-large  to small scales rather than the reverse-maps
to the  tilt of the spectrum being red or blue, while the speed of
the transition, parameterized by $\gamma$, is measurable  in the
tilt  (\ref{ns}).  Note  that the  scenario implies also the
consistency condition (\ref{consist})  which, given $\zeta$, is a
precise prediction.  Alternatively, the two  parameters $\gamma$
and  $\zeta$ are together  determined by  the tilt  and running of
the spectrum.  Our results
are applicable to a large class of similar models.


We can point out that exact scale-invariance is a natural prediction
of this model, without any fine tuning.  This is in fact not
completely ruled out by the data.  However, the data presently prefers
the model's ability to produce a slightly red spectrum, with a very
characteristic running, encoded in consistency condition
(\ref{consist}). This requires large critical exponents (of order
50); whether this is reasonable or not might be investigated in
explicit models of Phase I.  Finally, we note that fluctuations are
very nearly Gaussian~\cite{pog} and tensor modes are expected to be
negligible in our preliminary estimates.

We thank R. Brandenberger, 
O. Dreyer, S. Hossenfelder, M Joyce, L. Kofman, J. Khouri, and
F. Markopoulou for encouragement and discussion.  CRC thanks PI and
CITA for hospitality. This work was performed on the MacKenzie cluster
at CITA, funded by the Canada Foundation for Innovation. Research at
PI is supported in part by the Government of Canada through NSERC and
by the Province of Ontario through MEDT.


\begin{thebibliography}{99}

\bibitem{non-geometry}F. Markopoulou, hep-th/0604120, in {\it
Approaches to Quantum Gravity}, D. Oriti ed. CUP in press.
T. Konopka, F. Markopoulou and L. Smolin, hep-th/0611197.
\bibitem{Thiemann}T. Thiemann, Class.Quant.Grav. 15 (1998) 1463.
\bibitem{thooft}G. 't Hooft, gr-qc/9310006, hep-th/0003004.
\bibitem{susk}L. Susskind, J. Math. Phys. 36 (1995) 6377,
J. Maldacena,Adv. Theor. Math. Phys. 2 (1998) 231.
\bibitem{otherholo}L. Smolin, hep-th/0003056; R. Bousso, Rev. Mod. Phys. 74 (2002) 825-874.
\bibitem{liddle} A. Liddle and D Lyth, Cosmological Inflation and Large-scale structure,CUP, Cambridge 2000.
\bibitem{ks}H. Kodama and M. Sasaki, Prog. Th. Phys. 78 (1984) 1.
\bibitem{WMAP03cosmo} Spergel D.,  et~al., 2003, Astrophys. J. Suppl., 148, 175.
\bibitem{rob}
A. Nayeri, R. H. Brandenberger, C. Vafa, Phys. Rev. Lett. 97: 021302, 2006;
R. Brandenberger et al, hep-th/0608121 and hep-th/0608186; Biswas et al, hep-th/0610274.
\bibitem{renata} J. Ambjorn, J. Jurkiewicz, R. Loll, hep-th/0604212, in Approaches to Quantum Gravity", CUP; M. Reuter, F. Saueressig, Phys.Rev. D66 (2002) 125001.
\bibitem{Lewis:2000}A.  Lewis, A. Challinor,
A. Lasenby, Ap.J. 538,(2000), 473.
\bibitem{Lewis:2002ah}{Lewis}, A. \& {Bridle}, S. 2002, Phys. Rev. D, 66, 103511

\bibitem{data}G. {Hinshaw} et~al. astro-ph/0603451, {Readhead}
et~al. Ap.J 609 (2004) 498, Readhead et~al. Science, 306, 836;
{Sievers} et~al. 2005, astro-ph/0509203; Kuo et al., astro-ph/0611198;
{Halverson} et~al. Ap. J 568 (2002) 38; S. {Hanany} et~al. 2000,
Ap.J.Lett., 545, L5; {Hinshaw} et~al. astro-ph/0603451, {Leitch}
et~al. 2005, Ap. J., 624, 10; {Dickinson}, C., et~al. 2004, MNRAS,
353, 732; {Jones}, W.~C., et~al. 2006, Ap. J., 647, 823; {Piacentini},
F., et~al. 2006, Ap. J., 647, 833; {Montroy}, T.~E., et~al. 2006,
Ap. J., 647, 813; {Cole}, S., et~al. 2005, MNRAS, 362, 505; {Tegmark},
M., et~al. 2004, Phys. Rev. D, 69, 103501.
\bibitem{vsl0}J. Moffat, Int. J. Mod. Phys. D2 (1993) 351;A. Albrecht and
J. Magueijo, Phys.Rev. D59 (1999) 043516; J. Magueijo, Rept. Prog. Phys. 66, (2003) 2025.
\bibitem{pog}J. Magueijo and L. Pogosian, Phys. Rev. D67, 043518, 2003.
\bibitem{joy}A. Gabrielli, M. Joyce, F. Labini, astro-ph/0110451.
\bibitem{dav}D. Oaknin, hep-th/0305068; hep-th/0308078.


\end{thebibliography}
\end{document}